%% file: main.tex
\documentclass[pageno]{jpaper}
\usepackage[normalem]{ulem}

\usepackage{mathptmx} 
\usepackage{appendix}
\usepackage{fancyhdr}
\usepackage{enumitem}
\usepackage{multirow}
\usepackage[noabbrev,capitalise]{cleveref}
\newcommand{\ignore}[1]{}
\newcommand{\REVignore}[1]{}
\newcommand{\arxivIgnore}[1]{}
\newcommand{\arxivAdd}[1]{#1}
\usepackage{tikz}

\def\BibTeX{{\rm B\kern-.05em{\sc i\kern-.025em b}\kern-.08em
    T\kern-.1667em\lower.7ex\hbox{E}\kern-.125emX}}
\definecolor{aliceblue}{rgb}{0.94, 0.97, 1.0}
\usepackage{xcolor}

\usepackage{tcolorbox}
\tcbset{width=1\columnwidth,boxrule=1pt,colback=aliceblue,arc=1pt,left=2pt,right=2pt,boxsep=0.5pt}

\begin{document}

\definecolor{myblue}{rgb}{0, 0, 1}
\newcommand\revhl[1]{%
  \bgroup
  \hskip0pt\color{black!80!black}%
  #1%
  \egroup
} 

\arxivAdd{\author{Anish Saxena\thanks{The author can be reached at asaxena317@gatech.edu} \quad\quad\quad\quad Saurav Mathur\quad\quad\quad\quad Moinuddin Qureshi\\ \vspace{1mm}
Georgia Institute of Technology}}
\arxivAdd{\title{Randomized Line-to-Row Mapping for Low-Overhead Rowhammer Mitigations}}

\date{}

\maketitle

\thispagestyle{empty}
\input{./sections/0.abstract}
\input{./sections/1.introduction}
\input{./sections/2.background}

\input{./sections/4.methodology}
\input{./sections/3.policy}

\input{./sections/5.rubixd}
\input{./sections/6.related}
\input{./sections/7.conclusion}

\bibliographystyle{plain}
\bibliography{references}

\end{document}

%% file: sections/0.abstract.tex
\begin{abstract}
Modern systems mitigate Rowhammer using {\em victim refresh}, which refreshes the two neighbours of an aggressor row when it encounters a specified number of activations.
Unfortunately, complex attack patterns like Half-Double break victim-refresh, rendering current systems vulnerable. Instead, recently proposed secure Rowhammer mitigations rely on performing mitigative action on the aggressor rather than the victims. Such schemes employ mitigative actions such as {\em row-migration} or {\em access-control} and include AQUA, SRS, and Blockhammer. While these schemes incur only modest slowdowns at Rowhammer thresholds of few thousand, they incur prohibitive slowdowns (15\%-600\%) for lower thresholds that are likely in the near future. The goal of our paper is to make secure Rowhammer mitigations practical at such low thresholds.

Our paper provides the key insights that benign application encounter thousands of {\em hot rows} (receiving more activations than the threshold) due to the memory mapping, which places spatially proximate lines in the same row to maximize row-buffer hitrate.
Unfortunately, this causes row to receive activations for many frequently used lines. We propose {\em Rubix}, which breaks the spatial correlation in the line-to-row mapping by using an encrypted address to access the memory, reducing the likelihood of hot rows by 2 to 3 orders of magnitude. To aid row-buffer hits, Rubix randomizes a group of 1-4 lines. We also propose {\em Rubix-D}, which dynamically changes the line-to-row mapping. Rubix-D minimizes hot-rows and makes it much harder for an adversary to learn the spatial neighbourhood of a row. Rubix reduces the slowdown of AQUA (from 15\% to 1\%), SRS (from 60\% to 2\%), and Blockhammer (from 600\% to 3\%) while incurring a storage of less than 1 Kilobyte.

\end{abstract}

%% file: sections/1.introduction.tex
\section{Introduction}

\begin{figure*}
    \centering
     \includegraphics[width=2\columnwidth]{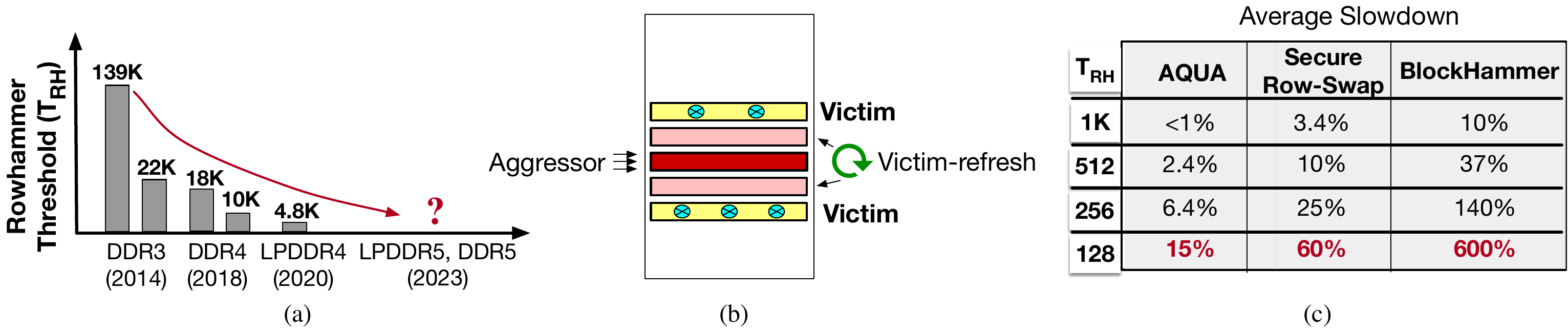}    
     \caption{(a) Trend of Rowhammer threshold (30x lower in 6 years) (b) Half-Double breaks victim-refresh. (c) Secure Rowhammer mitigations that are resilient against Half-Double, incur impractical slowdown at low thresholds ($T_{RH}$=128).} 
\label{fig:intro}
\end{figure*}

Rowhammer is a data-disturbance error where frequently activating a row induces bit flips in nearby rows~\cite{kim2014architectural}. Rowhammer is a severe security threat and has been used to leak confidential data and escalate privilege~\cite{seaborn2015exploiting, frigo2020trrespass, gruss2018another, kwong2020rambleed, aweke2016anvil, cojocar2019exploiting, gruss2016rowhammer, van2016drammer, jang2017sgx,kwong2020rambleed}. Rowhammer worsens with higher memory density. The number of activations  required to induce bit-flips, termed as the {\em Rowhammer Threshold ($T_{RH}$)}, has plummeted from 139K (DDR3) in 2014 to just 4.8K in 2020 (LPDDR4), as shown in \cref{fig:intro}~(a). The threshold is expected to reduce even further, and if the current trend continues (30X reduction in 6 years), we can expect $T_{RH}$ close to 100 by the end of this decade. Solutions that protect against Rowhammer must be viable not just at the current threshold, but also at future thresholds.

Rowhammer defenses typically incorporate a \textit{tracking mechanism} to keep activation counts for row and a \textit{mitigative action} that is performed when the activation count reaches the threshold.  The most popular form of mitigative action is {\em victim refresh}, which simply refreshes the nearby {\em victim rows} when the {\em aggressor row} reaches the specified activations. 
\revhl{Victim refresh has been deployed in commercial systems (e.g. DDR4, DDR5) in the form of TRR~\cite{frigo2020trrespass}.
However, the drastic reduction of $T_{RH}$ poses two problems.
First, due to severe area limitation in DRAM (up-to 9\% for per-row tracking~\cite{samsung_dsac}), TRR is unable to identify all aggressors, even in DDR5~\cite{samsung_dsac}. In fact, two recent whitepapers from JEDEC\cite{JEDEC-RH1, JEDEC-RH2} mention that ``in-DRAM mitigations cannot eliminate all forms of Rowhammer attacks".  
Second, even if tracking is perfect, the act of victim-refresh can itself be used to induce bit-flips. }
As shown in \cref{fig:intro}~(b), Half-Double~\cite{half-double} leverages victim-refresh to cause bit-flips at a distance-of-2 from the aggressor row, thereby breaking all defenses relying on victim-refresh. Thus, current systems remain vulnerable to Rowhammer.
\revhl{In this paper, we focus on mitigations resilient to complex patterns.}

Recent studies~\cite{aqua,srs,yauglikcci2021blockhammer} propose mitigations that are secure against complex attacks by performing the mitigative action on the aggressor instead of the victim (unlike victim-refresh).  AQUA~\cite{aqua} and SRS~\cite{srs} perform mitigation by migrating the aggressor row to another row, thereby breaking the spatial correlation between the aggressor and victim. Blockhammer~\cite{yauglikcci2021blockhammer} performs mitigation by limiting the number of activations to any row to less than $T_{RH}$, thereby preventing complex-pattern attacks that need a large number of activations to one row. While these schemes invoke high-overhead mitigating actions (row-migration incurs several microseconds and rate-control can increase latency by more than 10x), at current $T_{RH}$ only a small number of rows require any mitigation, and these schemes incur a modest slowdown.

Unfortunately, as $T_{RH}$ reduces, a larger number of rows reach the specified number of activations and hence would require mitigation, causing secure mitigation schemes to incur significant overheads.  \cref{fig:intro}~(c) shows the average slowdown from AQUA, SRS, and Blockhammer as the threshold is varied from 1K to 128. While these schemes have only a modest overhead at the 1K threshold, at a threshold of 128, AQUA suffers a slowdown of 15\%, SRS of 60\%, and Blockhammer of 600\%, rendering them impractical at lower thresholds. 

\textit{The goal of our paper is to make secure Rowhammer mitigations viable for practical adoption even at a low threshold (128), which is likely to be present by the end of this decade.}

\newpage
We define the rows that receive more than the threshold number of activations within  64ms as {\em hot rows}.  The key reason for slowdown at the lower threshold is the dramatic increase in the number of hot rows. On average, we observe only about 200 hot-rows with 512 or more activations, but 9500 hot-rows with 64 or more activations (45X more). Reducing the number of hot-rows reduces the slowdown from secure mitigations.

In this paper, we make the key observation that the primary cause of hot-rows is the memory mapping function, which determines the set of lines that co-reside within the same row. The activation count of a row is the sum of the activation counts caused by each of the 128 lines in an 8KB row. The memory-mapping in modern processors places lines with spatial proximity within the same row to maximize row-buffer hits. For example, Intel Coffee Lake~\cite{wang2020dramdig} mapping places the entire 4KB page within the same row and Intel Skylake~\cite{wang2020dramdig} round-robins the lines of each 4KB page between rows of two banks. Thus, 32-64 lines of each 4KB page co-reside within the same row. If the page is heavily accessed, then up to 64 lines would contribute to the aggregate activation count of the row. While each line incurs only a few activations within 64ms, the sum of activations due to all the lines makes the row a hot-row. Furthermore, typical workloads access only a small fraction of the memory within 64ms, like in our 16GB DDR4 system, where less than 5\% of rows are touched within 64ms.

Instead of concentrating activations to lines within hot-rows, spreading activations to the entire memory greatly reduces hot-rows. With this insight, we propose {\em Rubix}, a memory mapping that breaks spatial correlation of lines to rows by using an encrypted address to access memory. We present two flavors of Rubix: Static {\em (Rubix-S)} and Dynamic {\em (Rubix-D)}.

Rubix-S uses K-Cipher~\cite{kcipher}, a low-latency programmable bit-width cipher, for address-space randomization.  This cipher is kept in the memory controller.  When the memory controller services a memory access, it encrypts the line-address, accessing the memory with the encrypted line address 
Encryption randomizes the line-to-row mapping, so the lines co-resident in the same row have no spatial correlation. Thus, the likelihood that heavily accessed lines  get placed in the same row reduces significantly, virtually eliminating all the hot-rows (64 or more activations) for all of our workloads. This avoids invoking secure  mitigation actions and the resulting slowdown.

\newpage 

While line-address encryption minimizes  hot-rows, it also has virtually zero row-buffer hit rate. To improve the row-buffer hit rate, Rubix-S encrypts a {\em gang} of 1-4 contiguous lines. For example, with a gang size of 4, Rubix-S would not encrypt the bottom two least-significant bits of the line-address and only encrypt the remaining bits of the line-address.  With this organization, Rubix-S balances both row-buffer hit-rate and reduction in hot-rows.  Our evaluations show that at $T_{RH}$ of 128, Rubix-S reduces the slowdown of AQUA (from 15\% to 1\%), SRS (from 60\% to 3\%), and Blockhammer (from 600\% to 3\%) while requiring just 16 bytes of storage, thereby making it practical to deploy secure mitigations.

With Rubix-S, the group of lines that co-reside in the row are randomized, however, this group remains unchanged throughout the system uptime.  Rubix-D not only randomizes the line-to-row mapping but this mapping changes dynamically throughout the system runtime.  Instead of relying on encryption, Rubix-D uses an xor operation with a randomly generated key to perform randomization.  The mapping changes gradually from the current-key to the next-key.  Rubix-D performs the remapping vertically (gangs in the same position of different rows) instead of horizontally (gangs within the row), not only reduces hot-rows, but also making it much harder for an adversary to determine the set of rows that are spatially contiguous to each other (a critical step in launching a targetted Rowhammer attack). Our evaluations show that at $T_{RH}$ of 128, Rubix-D reduces the slowdown of AQUA (from 15\% to 1.5\%), SRS (from 60\% to 2\%), and Blockhammer (from 600\% to 3\%) while incurring a storage overhead of less than 1 KB.

\begin{figure*}
    \centering
     \includegraphics[width=2\columnwidth]{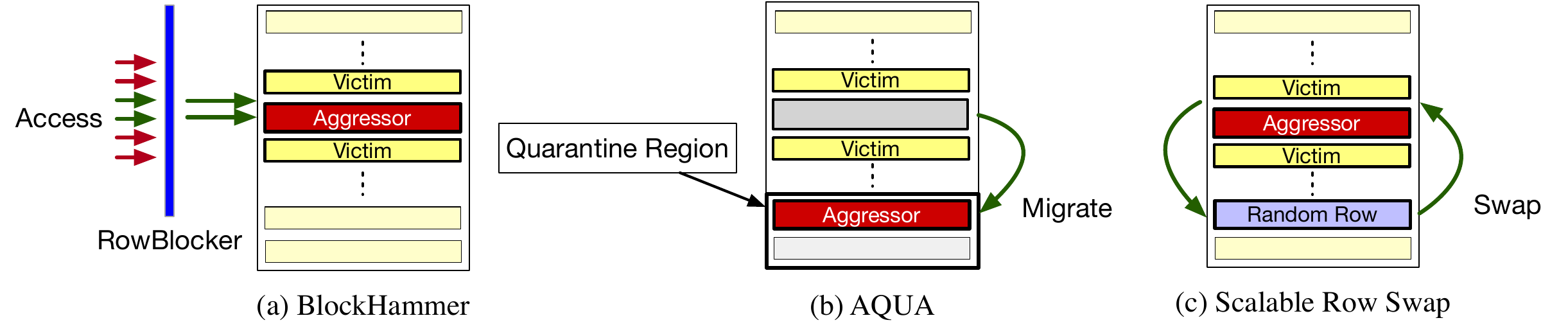}    
     \vspace{-0.05 in}
     \caption{Secure Rowhammer Mitigations: (a) BlockHammer controls rate of accesses to each row. (b) AQUA quarantines aggressor rows in a dedicated region. (c) Scalable Row-Swap (SRS) swaps the aggressor row with a random row.} 
         \vspace{-0.05 in}
\label{fig:motiv}
\end{figure*}

\vspace{0.05 in}
Overall, our paper makes the following contributions:
\begin{enumerate}
    \item To the best of our knowledge, this is the first paper to analyze the impact of memory (line-to-row) mapping on the efficacy of Rowhammer mitigations.  
   \item We demonstrate that the line-to-row mapping is the primary reason for hot-rows in benign workloads.
    \item We propose Rubix-S, which breaks the spatial correlation in line-to-row mapping by accessing the memory with an encrypted address (with gangs of 1-4 lines).

    \item We propose Rubix-D, which not only randomizes the line-to-row mapping but also changes this mapping continuously, making it much harder to identify rows that are in spatial proximity to each other.
\end{enumerate}

%% file: sections/2.background.tex
\section{Background and Motivation}

\subsection{Background on DRAM}\label{dram_org}

Modern DRAM-based memory is organized into several banks, each of which is a two-dimensional array of DRAM cells, organized as rows and columns. Each bank caches the most recently opened row in a {\em row buffer}. Data is accessed by bringing it into the row buffer.
To access data in another row, the bank clears the row buffer, followed by activation of the given row. DRAM cells leak charge and require periodic refresh operations (at 64ms). The parameter {\em $T_{RC}$ (Row Cycle Time)} determines the time between consecutive activations for a given bank. The $T_{RC}$ for current systems is about 45ns.

\subsection{Threat Model}
\label{threat_model}

We assume an unprivileged attacker that can run code on the system that is vulnerable to Rowhammer.
The attacker can run a process under user privilege and exploit Rowhammer to flip bits in critical data structures (such as page-table) or in the data of another program. We assume the Rowhammer bit-flip occurs at the victim location when any row in memory incurs more activations than $T_{RH}$ within the refresh interval of 64ms. Thus, the attack is successful if no mitigation is issued when a row has encountered more than $T_{RH}$ activations. 

\subsection{Memory Mapping}\label{memmap}

The memory-mapping function routes a given line address to a particular bank and row, thereby determining the set of lines that co-reside in a row~\cite{wang2020dramdig, heckel2023reverse}. It also affects row-buffer hit-rate and performance. Memory systems place spatially proximate lines in the same row and we consider two mappings used in Intel systems (our 8KB row buffer has 128 lines): 

\vspace{0.05 in}

\noindent{\bf Coffee Lake Mapping:} This mapping places consecutive 128 lines within the same row buffer. So, two consecutive 4KB pages would be resident in the same row. It uses a xor-based mapping hashed mapping for bank selection.

\vspace{0.05 in}

\noindent{\bf Skylake Mapping:} This mapping alternatively places a pair of lines between two banks (selected using xor).  So, for a 4KB page, lines 0,1,4,5 ... 60, 61 reside in a row of one bank, and lines 2,3,6,7 ... 62,63 are in row of the other bank. This mapping causes 32 lines from a 4KB page to reside in a row, with contents of four consecutive pages in the same row. 

\subsection{Rowhammer}
\label{row_hammer}

\revhl{Rowhammer is a data-disturbance error~\cite{rowhammer_crosstalk} where activating a row frequently induces bit-flips in nearby rows.}   The {\em Rowhammer Threshold ($T_{RH}$)} denotes the number of activations required on a row to induce bit-flips.  Rowhammer not only affects reliability but also system security, as the attacker can flip bits in the page table and take over the system. When  Rowhammer was \revhl{characterized} in 2014, $T_{RH}$ was 139K, whereas it reduced by 30x to 4.8K~\cite{kim2020revisiting} in 2020. \revhl{As memory gets denser, more nearby rows experience activations of the aggressor~\cite{lang2023blaster}.} Moreover, as $T_{RH}$ is likely to reduce further for future DRAM nodes, defenses for Rowhammer must  be designed not just for the current $T_{RH}$ but for future $T_{RH}$.

Hardware-based defenses for Rowhammer have two parts: activation-tracker and mitigating-action. \revhl{Several studies\cite{park_graphene:_2020, CBT, lee2019twice,kim2021mithril,MRLOC,PROHIT,qureshi2022hydra, bennett2021panopticon} have looked at storage-efficient trackers}. Comparatively, the mitigating action is less well studied. Most modern systems simply rely on victim-refresh, which is vulnerable to address-correlation attacks~\cite{HalfDouble}. Thus, modern systems continue to be vulnerable to Rowhammer attacks. 

\subsection{Secure Rowhammer Mitigation}

Performing the mitigating action on the aggressor row, instead of the victim, can prevent address-correlation attacks.  Recently several proposals have looked into such {\em Aggressor-focused mitigation}.  Figure~\ref{fig:motiv} shows three such schemes.

\vspace{0.05 in}

\noindent {\bf Blockhammer~\cite{yauglikcci2021blockhammer}} mitigates Rowhammer by controlling the access rate of frequently accessed rows, such that no row incurs more than $T_{RH}$ activations within 64ms, by delaying accesses for an appropriate time. As the adversary cannot perform an overwhelmingly large number of activations (required for Half-Double) on a single row, it prevents complex attacks.

\vspace{0.05 in}

\noindent {\bf AQUA~\cite{aqua}} mitigates Rowhammer by migrating the aggressor row when it receives $T_{RH}/2$ activations (halving of threshold due to tracker reset), to a {\em quarantine-region} in memory. AQUA breaks the spatial connection between aggressor and victim, limiting the time for an attacker to craft complex attacks.

\vspace{0.05 in}

\noindent {\bf Secure Row-Swap (SRS)~\cite{srs}} mitigates Rowhammer by swapping the aggressor row, once it has received $T_{RH}/3$ activations (reduction due to birthday-paradox attacks), with another randomly selected row in memory. Like AQUA, SRS breaks the spatial connection between the aggressor and the victim.

\subsection{Scalability Problem of Secure Mitigations}

While secure Rowhammer mitigations (such as Blockhammer, AQUA, and SRS) are resilient to complex attack patterns, they incur significantly more overhead than victim-refresh.
While performing two victim-refresh activation takes less than 100 nanoseconds, these schemes incur significantly more latency. For example, row migration required by both AQUA and SRS ties up the memory bus for several microseconds, during which the channel cannot service any requests. The problem is even worse for Blockhammer, where a given access can get delayed by several tens/hundreds of microseconds for rate control.

\begin{figure}[!htb]
    \centering
    \includegraphics[width=1\columnwidth]{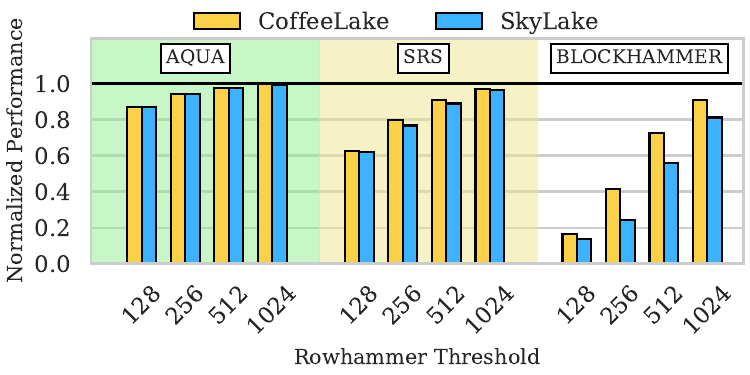}
    \caption{Normalized performance (versus no mitigation) of AQUA, SRS, Blockhammer as the threshold is varied. At $T_{RH}$ of 128, all schemes incur significant slowdowns.}
    \vspace{-0.05 in}
    \label{fig:secureperf}
   \vspace{-0.1 in}
\end{figure}

These schemes were designed and evaluated for the current thresholds of few thousand.  At such thresholds, very few rows reach the threshold in benign workloads, requiring mitigation.  
However, at lower thresholds, many more rows reach the threshold, which causes these high-overhead mitigations to occur more frequently, causing significant slowdowns.  Figure~\ref{fig:secureperf} shows the performance of AQUA, SRS, and Blockhammer as thresholds ranging from 1K to 128, for the Coffee Lake and Skylake memory mappings. The performance is normalized to the baseline which  uses Coffee Lake mapping.

We observe that at the threshold of 1K, AQUA and SRS both have a negligible slowdown, whereas Blockhammer suffers 10\% (Coffee Lake) to 25\% (Skylake).  However, at $T_{RH}$ of 128, all schemes incur increased overheads. AQUA incurs 15\% slowdown and SRS has 60\% slowdown. Blockhammer has 500\% to 600\% slowdown (note that 80\% reduction in normalized performance implies 5x slowdown).  These prohibitive overheads make secure mitigations impractical for adoption. 

\subsection{Goal of Our Paper}

The goal of our paper is to make secure Rowhammer mitigations viable at low thresholds ($T_{RH}$ of 128), which are likely to be present in the near future.  Ideally, we would like to accomplish this without incurring significant hardware overheads and develop a general framework that can be used even by other (future) secure Rowhammer mitigations as well. We discuss our methodology before describing our solutions.  

\arxivAdd{
\begin{figure*}
    \centering
     \includegraphics[width=1.7\columnwidth]{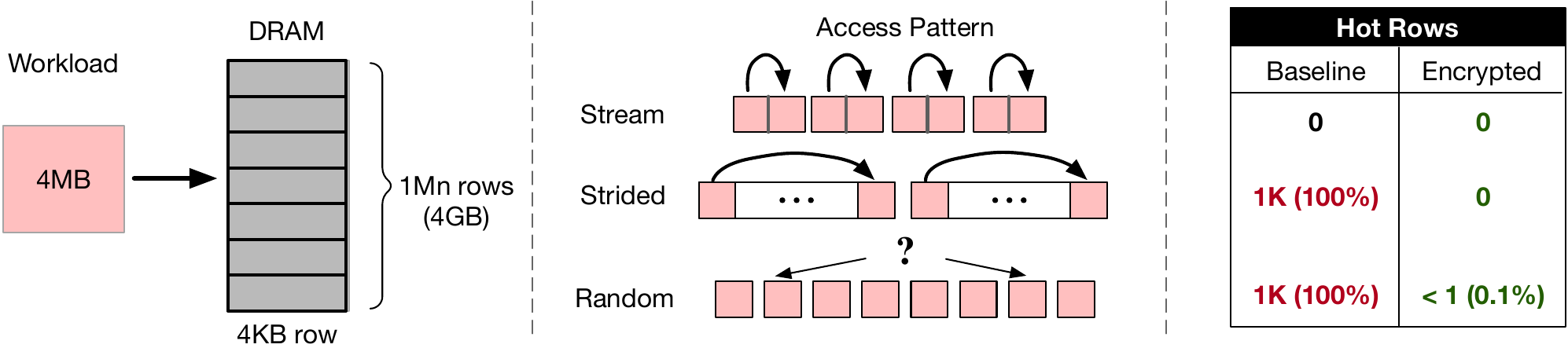}    
     \caption{Illustration: Understanding the impact of memory-mapping in forming hot-rows (a) System configuration (b) Workloads (c) Number of hot-rows for 4MB footprint (and 4KB rows).  Under baseline mapping, both stride-64 and random have 1K hot rows (100\%), however, with an encrypted address virtually all the hot-rows get eliminated.} 
\label{fig:synthetic}
\end{figure*}
}

%% file: sections/4.methodology.tex
\newpage
\section{Evaluation Methodology}

\subsection{System Configuration} 

We use the Gem5~\cite{lowe2020gem5} simulator to perform multi-core simulations in Syscall Emulation (SE) mode with an out-of-order core and DDR4 memory model.
We use DDR4 2400MT/s memory modelled using Micron MT40A2G4~\cite{micron_ddr4}.  Table~\ref{table:system_config} shows our baseline system configuration. \revhl{We use the open-adaptive memory page policy which keeps the row open for maximum of 16 accesses before closing it. Moreover, we use first-ready FCFS (FR-FCFS) scheduling policy to prioritize row hits and minimize unnecessary activations.}
We use the Coffee Lake mapping as our baseline. 
AQUA and SRS use the Misra-Gries~\cite{park2020graphene} tracker and for Blockhammer, we use an idealized SRAM tracker with one counter per row in memory. Due to tracker state reset, we use a tracker threshold of $T_{RH}/2$.

\begin {table}[h!]
\begin{small}
\begin{center} 
\vspace{-0.15 in}
\caption{Baseline System Configuration}
\begin{tabular}{|c|c|}
\hline
  Out-of-Order Cores           & 4 cores, 8 wide at 3GHz       \\ \hline
  Last Level Cache (Shared)    & 8MB, 16-Way, 64B lines \\ \hline
  Memory size                  & 16 GB -- DDR4 2400MT/s \\
  t$_{RCD}$-t$_{CL}$-t$_{RP}$-t$_{RC}$ & 14.2-14.2-14.2-45 ns\\
  Rows x Banks x Ranks x Channels     & 128K$\times$16$\times$1$\times$1 \\
  Size of row                  & 8KB  \\ \hline
\end{tabular}
\label{table:system_config}
\end{center}
\end{small}
\end{table}
\vspace{-0.15in}

\subsection{Workloads}

We evaluate with 18 SPEC2017~\cite{SPEC2017} \textit{rate} workloads and 16 mixed workloads (each with four random SPEC2017 workloads). We fast-forward 25 billion instructions and simulate for 250 million instructions. Table~\ref{table:wc} shows the Misses Per 1K Instructions (MPKI) and the average number of unique rows touched, and "hot-rows" with 64 or more activations (ACT-64+) and with 512 or more activations (ACT-512+).
\begin{table}[!htb]
  \centering
  \begin{footnotesize}
\vspace{-0.1 in}
  \caption{Workloads Characteristics: MPKI, Unique Rows Touched (within 64ms), and Hot-Rows (within 64ms).}
  \label{table:wc}
  \begin{tabular}{|c||c|c|c|c|}
    \hline
    & MPKI & {Unique Rows}& \multicolumn{2}{c|}{Total number of "Hot-Rows"} \\ \cline{4-5}
Workload	&	(LLC)	&	Activated	&	ACT-64+ & ACT-512+ \\ \hline \hline
blender   & 12.78 & 8.8K     & 347K  & 2.9K   \\
lbm       & 20.87 & 29.4K     & 70.3K & 0 \\
gcc       & 6.12  & 10.4K     & 21.8K & 384    \\
cactuBSSN    & 2.57  & 5.2K     & 12.2K  & 0    \\
mcf         & 5.81  & 4.9K     & 10.5K    & 425     \\
roms      & 3.33  & 27.9K     & 6.6K & 9     \\
perlbench & 0.71  & 11.4K        & 1.7K    & 0    \\
xz        & 0.40  & 10.8K      & 496    & 0    \\
nab       & 0.53  & 4.4K        & 189    & 0    \\
namd      & 0.37  & 3.4K        & 105    & 0    \\
imagick   & 0.13  & 1.1K        & 89    & 0    \\
bwaves    & 0.21  & 1.7K        & 20    & 0    \\
wrf       & 0.02  & 702        & 20    & 0    \\
exchange2 & 0.01  & 122        & 14   & 0    \\
deepsjeng & 0.25  & 68.1K        & 12    & 0    \\
povray    & 0.01  & 390        & 8    & 0    \\
parest    & 0.10  & 2.4K        & 3     & 0   \\
leela     & 0.02  & 879        & 0    & 0    \\

\hline \hline
Average & 3.01 &	10.7K &	9528 & 206 \\ \hline
  \end{tabular}
  \end{footnotesize}
\end{table}

%% file: sections/3.policy.tex
\section{A Case for Randomized Memory}

The reason secure Rowhammer mitigations incur significant overheads at low thresholds is because more rows reach the threshold number of activations (we refer to such rows as {\em hot-rows}). We identify the root cause of hot-rows to be the memory mapping function that determines the line-to-row mapping.  In this section, we first present this insight, then our workload characterization, then our solution {\em Rubix}, and results for slowdown and mitigation. 

\subsection{Dependence of "Hot Rows" on Mapping}

We illustrate the dependence of hot-rows on line-to-row mapping using a simple model, as shown in Figure~\ref{fig:synthetic} (a). The processor accesses a memory system containing one bank.  The memory system is 4GB and contains 1 million rows of 4KB each. We use sequential mapping that places the 4KB page within the same row.

We consider three kernels as shown in Figure~\ref{fig:synthetic} (b): \texttt{stream}, \texttt{stride-64}, and \texttt{random}, where each kernel has a footprint of 4MB and makes 1 million accesses to the memory within 64ms. We deem a row to be a hot-row if it has at least 64 activations. We analyze the number of hot-rows for these kernels.

For the \texttt{stream} kernel, the first access causes an activation, and subsequent 63 accesses all get a row-buffer hit. Therefore,  a million memory accesses cause a total of only 15.6K activations, which get spread equally over the 1K rows, with a uniform activation rate of about 16 activations per row, with no hot-rows.
The \texttt{stride-64} kernel has a stride of 64 lines and each access goes to a different page. When all pages are exhausted, the stride continues with the next line on the page. As each memory access causes an activation, this kernel incurs 1 million activations, spread equally over 1K pages, and each row gets 1K activations. Thus, all the 1K rows are hot-rows.
The \texttt{random} kernel accesses a random line in memory.  The likelihood of a row buffer hit is negligibly small, so the 1 million accesses cause 1 million activations, spread over 1K rows. The average number of activations per row are 1000 (standard deviation of 32), with more than 99\% of the rows having more than 900 activations. Thus, we deem all the 1K rows to be hot-rows. The results are summarized in Figure~\ref{fig:synthetic} (c). 

The conventional mapping of placing sequential lines in the same row buffer is the root cause of hot-rows for both the stride pattern and the random pattern. We have 64 lines that cause activation of the same row in memory, thus compounding the total number of activations incurred by the given row.  

Consider an alternative mapping that uses an encrypted line-address to access the memory system. 
There are 64K lines in a 4MB footprint. These 64K lines would be spread over the 1 million rows in memory. We estimate (using binomial distribution) that 61.5K rows have exactly 1 line from the kernel mapped to them, 1.9K rows with 2 lines, and 40 rows with 3 lines (no row with 4 or more lines). 
For both \texttt{stream} and \texttt{stride}, each line gets accessed 16 times.  So, we have 61.5K rows with 16 activations, 1.9K rows with 32 activations, and 40 rows with 48 activations. Thus, no row is deemed a hot-row. For \texttt{random}, we estimate the expected number of hot-rows to be 0.4, so less than 1 row will be deemed a hot-row. Thus, randomizing the line-to-row mapping eliminates the hot rows of all three kernels.

\subsection{Characterizing Lines in Hot-Rows}

For our baseline system, we want to understand how many lines (out of the 128 lines) of the row contribute to making the row a hot-row.  For each row that reaches 64 activations, we measure the number of lines in the row that encountered at least 1 activation. Table~\ref{table:hot} shows the percentage of hot-rows that had 1-8 lines, 8-16 lines, 32-64 lines, and 64-128 lines (and the average) contributing to the row activation counts.

\begin{table}[!htb]
  \centering
  \begin{footnotesize}
\vspace{-0.1 in}
  \caption{Number of lines that add to activation counts of hot-rows (data for workloads with 100+ hot-rows).}
   \vspace{0.05 in}
  \label{table:hot}
  \begin{tabular}{|c||c|c|c|c|}
    \hline
    & \multicolumn{4}{c|}{Number of Activating Lines in a Hot-Row} \\ \cline{2-5}
Workload	&	1-32 & 32-64 & 64-128 & Average\\ \hline \hline
blender   & 2\%  & 98\%  & 0 & 60\\
lbm       & 0 & 100\% & 0 & 58\\
gcc       & 1\% & 99\%  & 0 & 60 \\
cactuBSSN   & 0 & 100\% & 0 & 63   \\
mcf         & 0 & 100\% & 0 & 52    \\
roms      & 3\% & 97\% & 0 & 51    \\
perlbench & 7.3\%    & 92\% & 0  & 47  \\
xz        & 0 & 100\% & 0  & 57  \\
nab       & 0 & 100\% & 0  & 58  \\
namd      & 0 & 100\% & 0  & 54  \\

\hline \hline
Average & 2\% & 98\% & 0 & 56\\ \hline
  \end{tabular}
  \end{footnotesize}
\end{table}

We observe that for 98\% of hot-row activations come from at-least 32 lines in the row.  
On average, 56 out of 128 lines incur at least one-activations within the hot-row.  This validates our hypothesis that hot-rows occur because many lines of the row contribute to the activation counts. Thus, the line-to-row mapping which decides which set of lines co-reside within the same row is the main reason for the occurrence of hot-rows.

\subsection{Rubix: Randomized Line-to-Row Mapping}

\begin{figure}[!htb]
    \centering
    \vspace{0.1 in}
    
    \includegraphics[width=0.8\columnwidth]{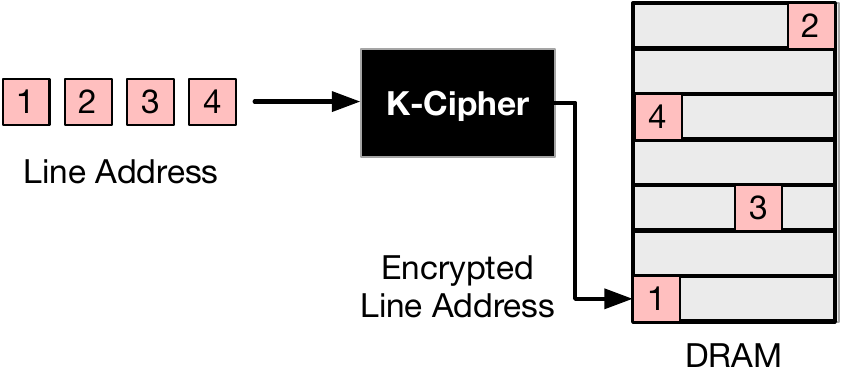}
    \caption{Overview of Rubix-S: breaking spatial correlation in line-to-row mapping with encrypted line-address.}
    \label{fig:rubix_line_level}
\end{figure}

{\em Rubix} breaks the spatial correlation of lines to row by using an encrypted address to access memory.
Figure~\ref{fig:rubix_line_level} shows an overview of the static version of Rubix, called {\em Rubix-S}. Consider the access pattern where requests for four consecutive lines A, B, C, D are set to memory. In conventional mapping, these four lines will co-reside within the same row.  However, with encryption, these lines get scattered to different rows.

Rubix-S uses K-Cipher~\cite{kcipher}, a low-latency programmable bit-width cipher, for address-space randomization.  \revhl{K-Cipher is kept in the memory controller and incurs a latency of 3 cycles (with 10nm process technology~\cite{kcipher})}.  On a memory access, it encrypts the line-address which is used to access the memory (as we have 16GB memory, we use a 28-bit cipher). Encryption randomizes the line-to-row mapping, breaking the spatial correlation between lines that co-reside in the row. 

The exact line-to-row mapping depends on the 96-bit key of the K-Cipher. The key is set to a random value (based on PRNG) at boot time.  As each system will have a different key, the memory mapping for each system will be different.

\subsection{Recouping Row-Buffer Hits via Gangs}

\begin{figure}[!htb]
    \centering
    \includegraphics[width=0.9\columnwidth]{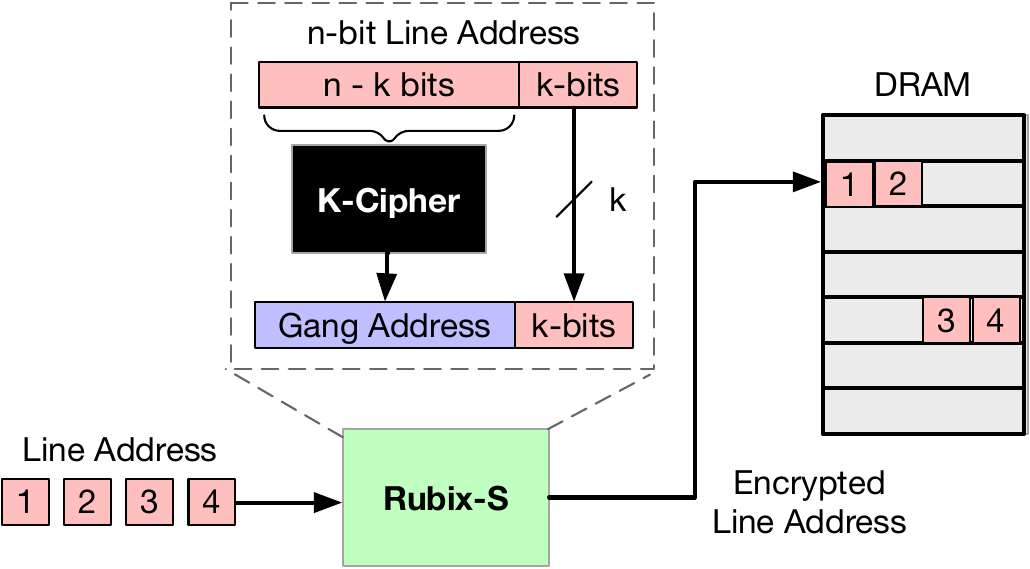}
    \caption{Rubix-S: Using gang-address encryption to balance both row-buffer hits and reduced hot-rows .}
    \label{fig:rubix_gang_level}
 \vspace{-0.05 in}
\end{figure}

While line-address encryption virtually eliminates hot-rows, it degrades the row-buffer hit-rate to approximately zero. Rubix minimizes hot-rows while still retaining some row-buffer hits, by encrypting a {\em gang} of 2-4 contiguous lines. Figure~\ref{fig:rubix_gang_level} shows Rubix-S with gang-level randomization. 

Instead of encrypting the entire n-bit line-address, Rubix-S skips the $k$ least significant bits of the line-address and only encrypts the {\em gang-address}, which is the remaining (n-k) bits. The encrypted gang-address is then concatenated with the unmodified k-bits, and this line-address is used to access the memory.
\revhl{Thus, lines in a gang co-reside in the same row, providing temporal locality which aids row-buffer hits.} For example, in Figure~\ref{fig:rubix_gang_level}, lines 1 and 2 co-reside in the same row.  Note that with k-bits, we would have a gang-size of $2^k$ lines and if k is set to zero, this design degenerates into Rubix-S with line-address encryption. We  denote Rubix-S with a gang-size of $X$ lines as {\em Rubix-S (GSX)}. The size of the cipher is adjusted per gang size, so Rubix-S (GS4) uses a 26-bit K-cipher.

\begin{figure*}[!htb]
    \centering
    \includegraphics[width=1\textwidth]{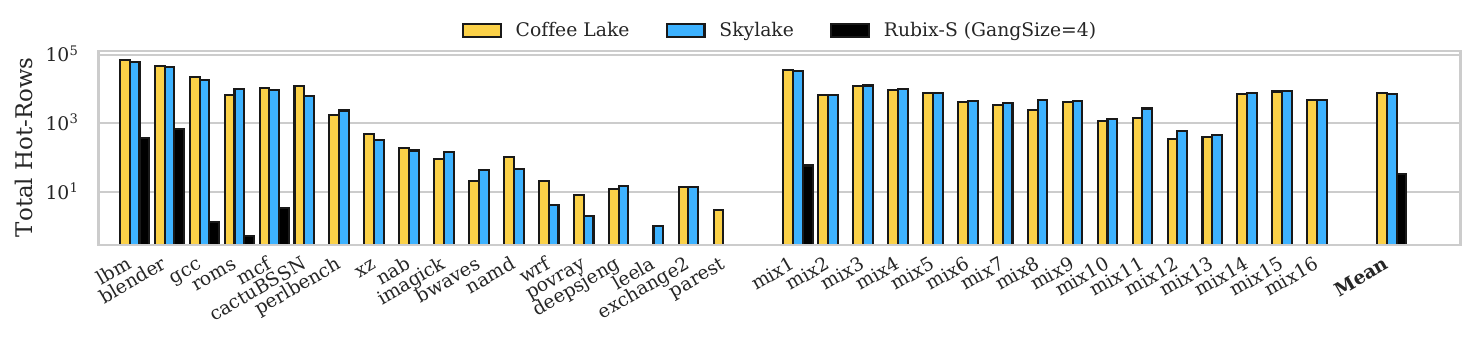}
   \vspace{-0.2 in}
    \caption{Number of hot-rows (activations of 64 or more) with Coffee Lake mapping, Skylake mapping, and Rubix-S with Gang-Size of 4 (GS4).  While baseline mappings have more than 7,000 hot rows on average, Rubix-S (GS4) reduces it by 220x to 33. 
    }
    \label{fig:hotrows}
  \vspace{-0.1 in}
\end{figure*}

\begin{figure*}[!htb]
    \centering
    \includegraphics[width=0.95\textwidth]{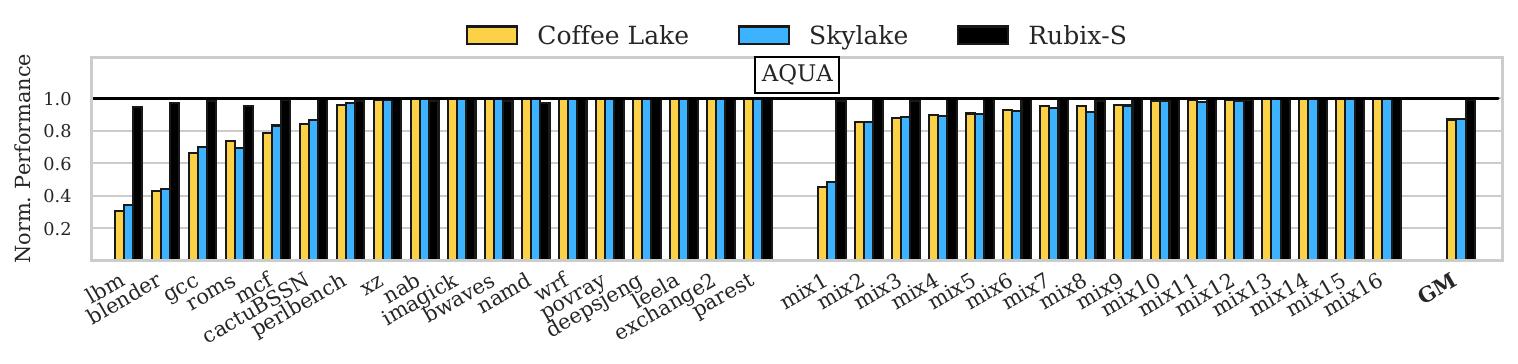}
   \vspace{-0.1 in}
    \label{fig:aqua_rubix_s}

    \centering
    \includegraphics[width=0.95\textwidth]{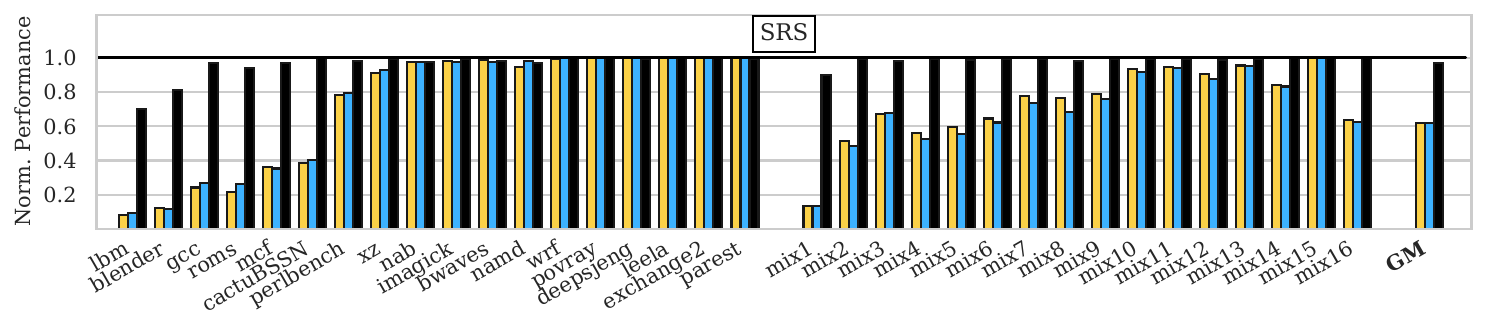}
   \vspace{-0.1 in}
    \label{fig:srs_rubix_s}
    \centering
    \includegraphics[width=0.95\textwidth]{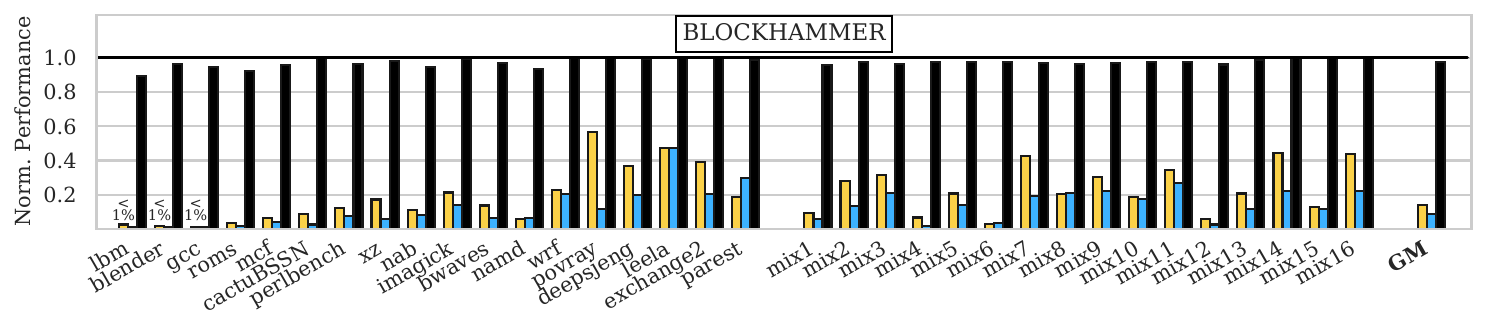}
    \label{fig:bh_rubix_s}
 \caption{Performance of secure mitigations at $T_{RH}$ of 128 for Intel mappings and Rubix-S, normalized to an unprotected Coffee Lake baseline. Rubix-S uses GS4 for AQUA and SRS, and GS1 for Blockhammer, and reduces the 
  average slowdown to 1.1\%, 3.1\%, and 2.9\%, respectively (down from 15\%, 60\%, and 600\%), making them viable at ultra-low thresholds.
 }
    \label{fig:rubix_s_perf}
\end{figure*}

\subsection{Results: Impact on Mitigations}

We observe that Rubix-S (GS1, line-level) eliminates all the hot-rows for our workloads. Figure~\ref{fig:hotrows} shows the number of hot-rows for the baseline system with Coffee Lake mapping, Skylake mapping, and Rubix-S (GS4).  
Rubix-S eliminates hot-rows for all but six workloads.
On average, Coffee Lake and Skylake mappings have 7.6K and 7.2K hot-rows respectively, whereas Rubix-S (GS4) reduces it by 220x to only 33. \revhl{Line-to-row mapping is a significant determinant of hot-rows, and our design significantly reduces hot-rows. 
Mitigations are invoked much less, greatly reducing the performance overheads.
}

\subsection{Results: Impact on Performance}

\cref{fig:rubix_s_perf} shows the performance of AQUA, SRS, and Blockhammer with Intel Coffee Lake, Skylake, and Rubix-S mappings. Performance is  normalized to an unprotected Coffee Lake baseline. Both the Intel mappings incur unacceptable slowdown with secure mitigations. We compare Rubix-S with Coffee Lake mapping which performs slightly better than Skylake. Coffee Lake incurs a significant average slowdown of 15\% for AQUA, while Rubix-S reduces it to a negligible 1\% (for gang-size 4). SRS and BlockHammer are impractical with baseline policies, incurring 60\% and 600\% average slowdown, respectively. Rubix-S not only enables SRS and BlockHammer with a negligible average slowdown of 3.1\% (GS 4) and 2.8\% (GS 1), respectively, it retains application-level performance with a worst-case slowdown of 42\% for lbm with SRS and just 11\% for BlockHammer, 28X and 350X improvement. 

Overall, Rubix-S makes secure mitigations viable even at ultra-low $T_{RH}$ of 128 with just 2-3\% overhead. 
While we do not change access scheduling and DRAM page policies, fine-tuning them would likely reduce the overheads even further.

\subsection{Results: Sensitivity to Gang-Size}

\begin{figure}[!htb]
    \centering
    \vspace{-0.1in}
    \includegraphics[width=0.9\columnwidth]{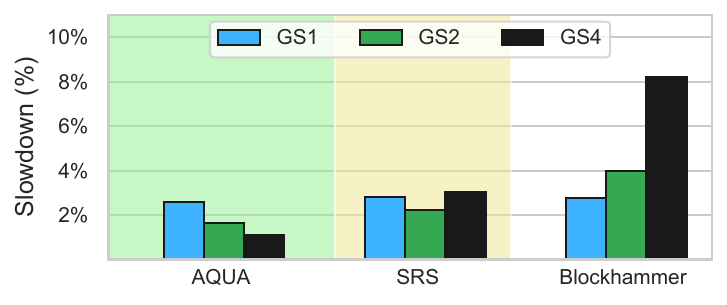}
    \vspace{-0.1in}
    \caption{Performance of Rubix-S with Gang-Size of 1-4.}
    \vspace{-0.15 in}
    \label{fig:rubix_s_gs}
\end{figure}

Gang-size (GS) balances row-buffer hits and reduction in hot-rows. With larger GS, row-buffer hit rate increases along with hot-rows and mitigation overheads.  \cref{fig:rubix_s_gs} shows the performance of secure mitigations with Rubix-S as GS is varied from 1 to 4. 
Due to high mitigation overhead, Blockhammer works best with GS1 which eliminates hot-rows. AQUA has lower overhead mitigation, and GS4 works best which retains row-buffer hits. For SRS, GS2 offers the best balance between row-buffer hits and minimizing hot-rows. Thus, the best GS size depends on the scheme and the mitigation overhead.

\begin{figure*}
    \centering
     \includegraphics[width=2\columnwidth]{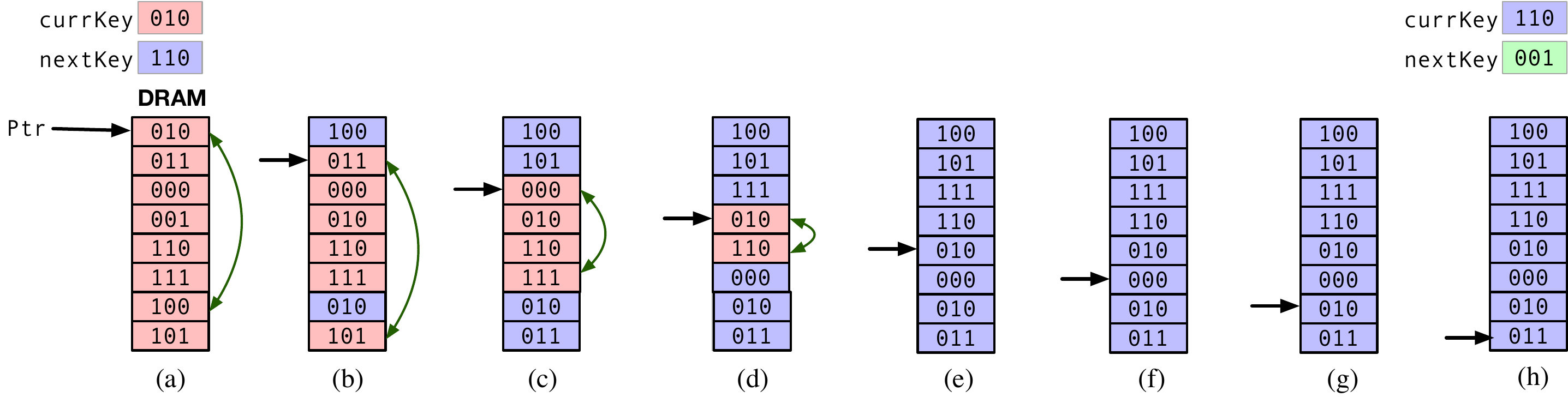}    
     \vspace{-0.05 in}
     \caption{An example of dynamically changing xor-based mapping. The effective address is the line-addressed xor-ed with a key. The dynamic remapping algorithm gradually remaps all the lines from a currKey (010) to the nextKey (110). } 
       \arxivAdd{\vspace{-0.1 in}}
\label{fig:xormap}
\end{figure*}

\subsection{Results: Impact on Row-Buffer Hits}

A key effect of small gang-size is decreased row-buffer hit rate. The baseline Coffee Lake and Skylake policies provide an average row-buffer hit rate of 55\% and 63\%, respectively. Rubix-S shows a gradual increase in row-buffer hit-rate from 0 with GS1, to 19\% at GS2 to 31\% at GS4, with up-to 2.7X more activations for GS1.
Thus, GS2 and GS4 recoup some of the row-buffer hits. 
The overall system performance depends not only on row-buffer hits but also on mitigation overheads.

\subsection{Results: Storage and Power Overheads}

Rubix requires negligible power for the K-Cipher and address mapping logic. The major overhead is from increased DIMM power due to lower row-buffer hit rate. We use Micron's power calculator~\cite{micron:calc} to compute DRAM power. Rubix-S increases the DRAM power by 120mW at a gang-size of 4 (a 4.3\% increase), and by 300mW at gang-size of 1 (10.6\% increase), due to a lower row-buffer hit rate than baseline that result in additional activations. The power consumption of Rubix-S with secure mitigations remains within 10\% of the baseline, because of virtually eliminating mitigations, unlike existing memory mappings, which incur prohibitive energy overheads.

\subsection{Security Analysis of Rubix-S}
\label{sec:rubixs-security}

The security of Rubix-S stems from the security of the underlying mitigation schemes (SRS, AQUA, Blockhammer). The security guarantees of these schemes are not dependent on using a specific memory-mapping. Rubix-S remains secure because we simply change the memory mapping. 

\subsubsection{Defining TRH}
We define \emph{$T_{RH}$} as the minimum number of activations to {\bf at least} one row within 64ms which causes a bit flip via any attack pattern (single-sided, double-sided, many-sided or Half-Double\cite{HalfDouble} or a future attack pattern). So to ensure security of our solution, our only assumption is:

\begin{tcolorbox}
A successful Rowhammer attack requires activating {\bf at least} one row more than \emph{$T_{RH}$} times within a refresh period.
\end{tcolorbox}

\subsubsection{Security of SRS, AQUA, and BlockHammer}

SRS and AQUA rely on row migration to guarantee that now row receives more than $T_{RH}$ activations within a 64ms window. SRS does so by randomization, guaranteeing that even under continuous attacks for several years, the likelihood of randomly finding migrated rows is negligibly small. With AQUA, a row that receives $T_{RH}/2$ activations is moved to a quarantine region, and by design, it guarantees that no physical row will ever receive more than 
$T_{RH}$ activations.  The security of BlockHammer depends on rate control. BlockHammer controls the activation rates to a physical row such that no row ever receives more than $T_{RH}$ activations. These schemes rely on accurate tracking of row counts, and we use Misra-Gries tracker for SRS and AQUA, and one-counter-per-row for BlockHammer, which provide guaranteed tracking. The security guarantees of SRS, AQUA, Blockhammer are applicable for all access patterns (including Half Double) and all possible memory mapping (the mapping of lines to rows).

\subsubsection{Proving  Security of Rubix-S Using Lemmas}

Rubix-S reduces performance overhead of secure mitigations, while retaining the security guarantees of the underlying SRS, AQUA, and Blockhammer schemes, which are secure against all access patterns and work with any memory mapping.

\begin{tcolorbox}[boxrule=1pt,left=5pt,right=5pt,top=1.5pt,bottom=1.5pt]
\textbf{Lemma-1:} \textit{The security guarantee of SRS, AQUA, and Blockhammer is not dependent on memory mapping, so these designs are secure for {\bf all} memory mappings. }
\end{tcolorbox}

\begin{tcolorbox}[boxrule=1pt,left=5pt,right=5pt,top=1.5pt,bottom=1.5pt]
\textbf{Lemma-2:} \textit{Rubix-S is a memory mapping which randomizes the line-to-row mapping.}
\end{tcolorbox}

From Lemma-1 and Lemma-2, it follows that secure mitigations continue to be secure with Rubix-S.  
For example, Half Double requires that an aggressor row be activated about 100x more times than $T_{RH}$. As no row is activated $T_{RH}$ times, all schemes are secure against Half-Double, even with Rubix-S.

\arxivAdd{
\vspace{-0.5mm}
\subsubsection{Performance for Worst-Case Patterns}

Rubix-S reduces the slowdown for typical applications, not for the worst-case.  
For example, an attacker can Flush+Reload~\cite{yarom2014flush+} the same cache line to cause many activations on a row, incurring significant slowdown even in the presence of Rubix. 
A motivated adversary can also learn the set of lines that map to the same row using timing attacks to cause hot-rows and high slowdown. However, these attacks are also possible in baseline and don't impact security of Rubix. 
}

%% file: sections/5.rubixd.tex
\newpage
\section{Rubix-D: Dynamic Randomization}

With Rubix-S, lines that co-reside in the row is randomized, however, this mapping remains unchanged throughout the system uptime.  In this section, we propose {\em Rubix-D}, an alternative approach that not only randomizes the line-to-row mapping but changes this mapping dynamically throughout the system runtime. 
Rubix-D reduces hot-rows and makes it much harder to determine rows that are spatially contiguous to each other (a critical step in a targetted Rowhammer attack).

We adapt the ideas presented in seminal works on dynamic memory remapping~\cite{start-gap,securityrefresh} 
to suit our constraints and objectives.
Rubix-D uses an xor operation with a randomly generated key to perform randomization~\cite{securityrefresh}.  The mapping is changed gradually from a given key to a new key.  In this section, we first provide an example of dynamically changing xor mapping, then present Rubix-D, and finally the results.

\subsection{Overview of Xor-Based Remapping}

Figure~\ref{fig:xormap} provides an example of the xor-based dynamic remapping for a memory containing 8 lines (000-111). The system contains a pointer (Ptr) to aid with remapping and two sets of keys currKey and nextKey.  The effective line-addresss is computed as the xor operation with one of the keys.  At the start of the remapping epoch, all lines use the currKey whereas by the end of the epoch, all lines use the nextKey.  We perform remapping every 100 accesses. Figure~\ref{fig:xormap} (a) shows the mapping at the start of the epoch with all lines located at their original address xor-ed with the currKey (010). After 100 accesses, the   first remapping is invoked, so the physical location 000 (pointed by the Ptr) is swapped with the destination 110 (Ptr xor-ed with the nextKey). Ptr is incremented to 001.

The next three remappings (every 100 accesses) also result in swaps (Figure~\ref{fig:xormap} (b), (c), and (d)) and the pointer is incremented accordingly.
For the next four remapping episodes, the swap operation is skipped as the Ptr points to an already remapped line. 
After 8 episodes, all lines use the mapping with nextKey, as shown in Figure~\ref{fig:xormap}(h). At this point, the currKey is revised to currkey xor-ed with nextKey, and the nextKey is initialized to a new value obtained using a hardware-based PRNG. The Ptr is reset to 000, indicating a new epoch.

We translate line-address to physical-address in two steps:

\vspace{0.05 in}
\noindent {\bf (1)} Translate line-address $L$ to $L'$ = (L xor currKey). 

\vspace{0.05 in}
\noindent {\bf (2)} Perform two checks: First, is $L'$ $<$ $Ptr$? and Second, is ($L'$ xor nextKey) $<$ $Ptr$?.  If either is yes, $L'$ = ($L'$ xor nextKey).  
\vspace{0.05 in}

The memory access is thus routed to location $L'$. 
The simple xor and checks operations are performed within one cycle.
Thus, xor-based dynamic remapping randomizes line-to-row addresses with negligible SRAM (three registers -- currKey, nextKey, and Ptr)) and latency (one cycle). For properties and proof of xor-based randomization, please refer to ~\cite{securityrefresh}.

\subsection{Pitfall of Xor at Randomizing Line-to-Row}

While xor-based mapping dynamically randomizes memory addresses, we cannot directly apply it in our context, due to the linear mapping of xor. 
For example, if there are 128 lines co-residing in a row, then after an xor with a random key, these 128 lines still co-reside in one row (at another location). As all the top (n-7) bits of the lines that get mapped to the same row are identical, an xor with the (n-7) bits in the key results in the same remapped value. 
Reordering of lines within the destination row, unfortunately, does not reduce the likelihood of it becoming a hot-row.
Instead, our proposal {\em Rubix-D} reorganizes the xor-based mapping to dynamically randomizes the group of lines that co-reside in a row.

\begin{figure*}[!htb]
    \centering
    \includegraphics[width=0.95\textwidth]{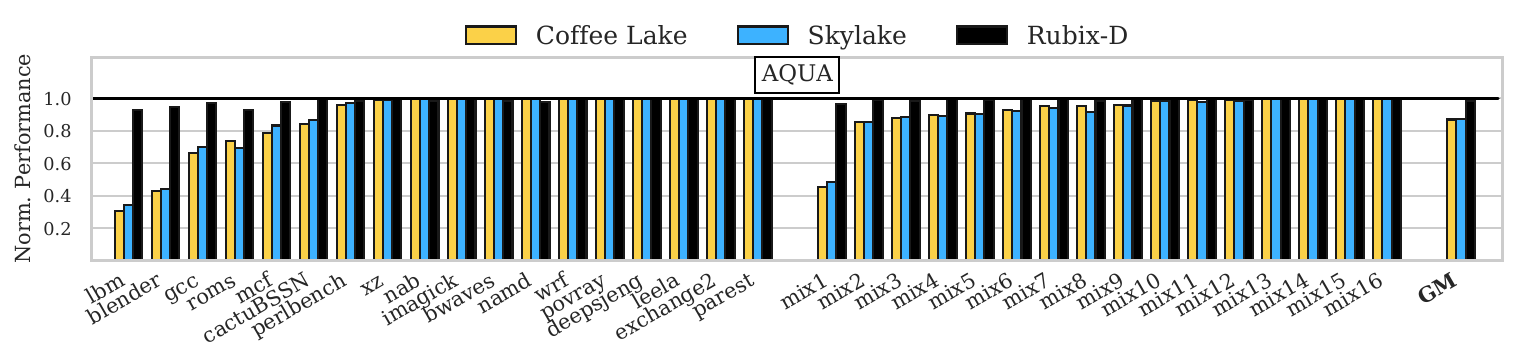}
   \vspace{-0.1 in}
    \label{fig:aqua_rubix_d}
    \centering
    \includegraphics[width=0.95\textwidth]{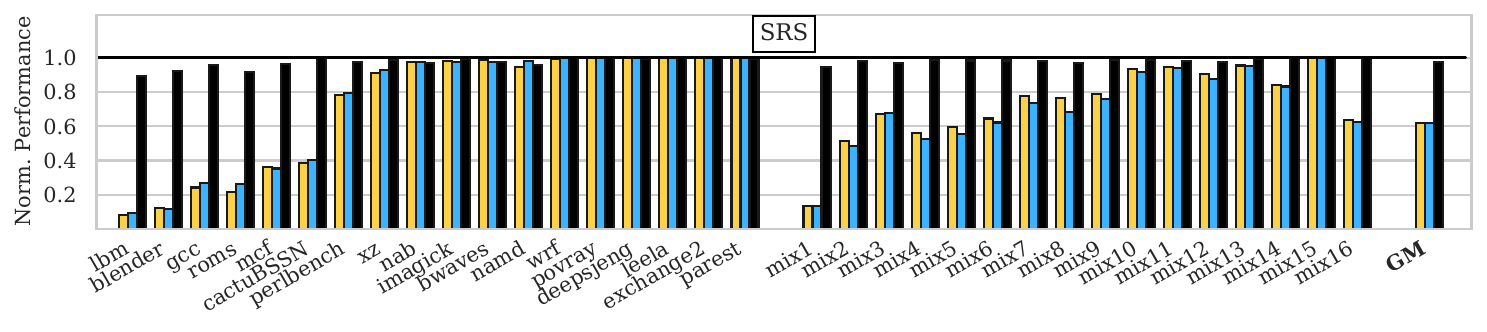}
   \vspace{-0.1 in}
    \label{fig:srs_rubix_d}
    \centering
    \includegraphics[width=0.95\textwidth]{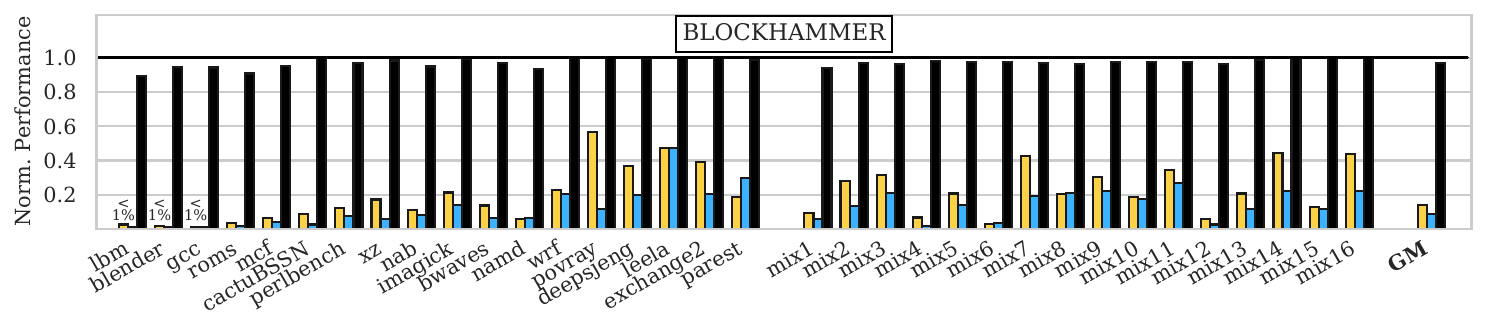}
    \label{fig:bh_rubix_s}
 \caption{Performance of secure mitigations at $T_{RH}$ of 128 for Intel mappings and Rubix-D, normalized to an unprotected Coffee Lake baseline. With GS4 for AQUA, GS2 for SRS, and GS1 for BlockHammer, Rubix-D incurs a low average slowdown of 1.5\%, 2.3\%, and 2.8\%, respectively (down from 15\%, 60\%, and 600\%).
 }
    \label{fig:rubix_d_perf}
\end{figure*}

\subsection{Overview of Rubix-D}

\begin{figure}[!htb]
    \centering
    \includegraphics[width=0.9\columnwidth]{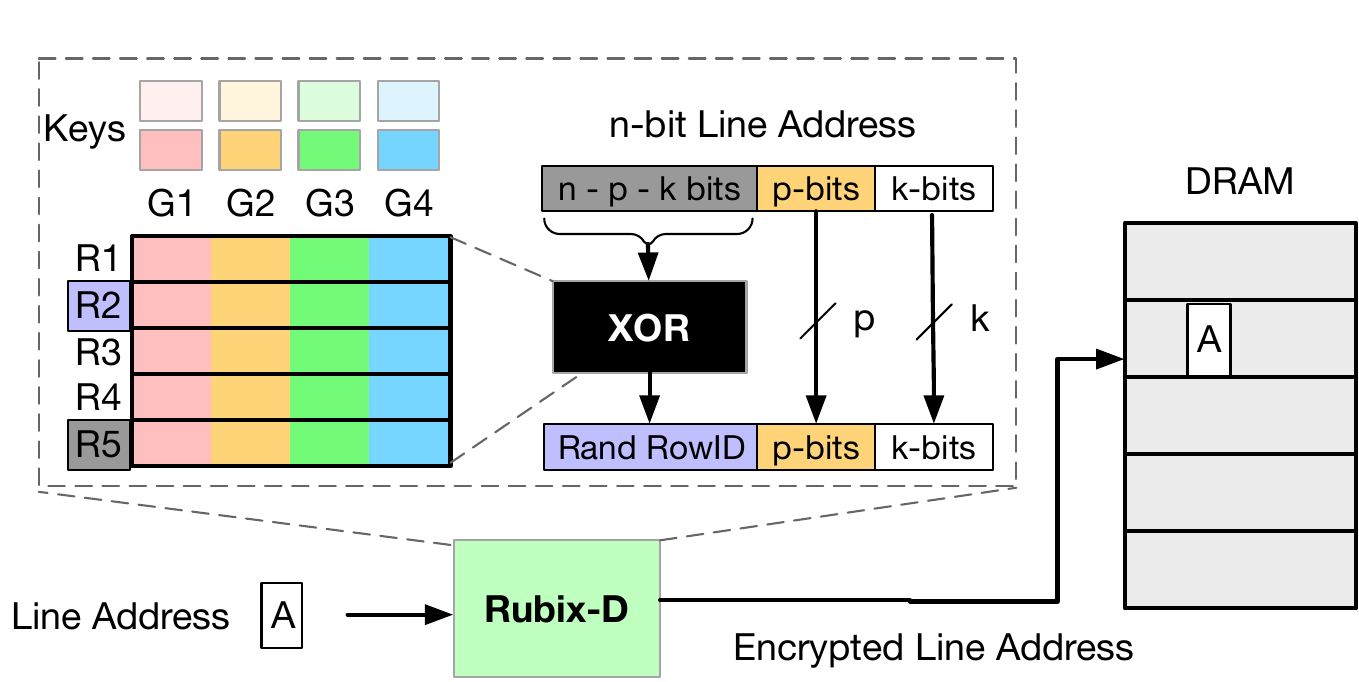}
    \caption{Overview of Rubix-D.  Each gang is independently routed to a random row, so all the gangs of a row get scattered in memory.}
    \label{fig:rubixd}   \vspace{-0.15 in}
\end{figure}

Figure~\ref{fig:rubixd} shows an overview of Rubix-D.  We randomize gangs vertically (across rows but for same gang-in-row). For $G$ gangs in a row, we provision $G$ sets of remapping circuits (currKey, nextKey, and Ptr). 
As each gang in the row uses a different key, gangs co-residing in the same row in the baseline are scattered to different rows in memory, breaking the spatial correlation between gang mapping to a row. 

In Figure~\ref{fig:rubixd}, the memory has 4 gangs in a row (colored red, yellow, green, blue). The same-colored gangs across all the $R$ rows of memory form a {\em vertical-group (v-group)}. Each v-group is allocated a pair of keys (curr and next) and a pointer.  At the memory controller, the line-address is split into three parts: the least significant $k$ bits identify the line-in-gang, next $p$ bits identify the gang-in-row, and remaining $n-p-k$ bits identify the row-address. 
Rubix-D keeps the $k+p$ bits of the line address unchanged, randomizing only the bits for (global) row address.  The $p$ bits identify the v-group and that v-group's currKey, next-Key, and Ptr translate the row-address to the remapped-row-address. The remapped-row-address is concatenated with the $p+k$ bits to form the remapped-line-address, which is used to access the memory.

With a 28-bit line address (16GB memory), Rubix-D with gang-size of 4 lines uses 2 bits to identify line-in-gang, the next 5 bits for gang-in-row, and the remaining 21 bits for global row address. 
With less than 8 bytes for each pair of keys and ptr, we need total SRAM of 512 bytes (for 32 v-groups).

\subsection{Remapping Rate and Remapping Period}

\revhl{The {\em Remapping-Rate (RR)} determines the frequency of remapping.  We set RR to occur with 1\% probability on each activation.  Thus, v-gangs with more activations are remapped more frequently.  During remap, the gangs pointed by the Ptr of the v-group are swapped with their destination (based on the nextKey). 
At gang-size of 4, 4 lines are streamed from source and destination rows and their content is swapped, incurring 3 activations, 8 CAS reads, and 8 CAS writes.
Half of the remap operations are skipped (as those gangs are already remapped), so the swap incurs 1.5 activations on average, which for an RR of 1\% is an average overhead of 1.5\% activations.
Thus, remap operations incur low average overheads.}

The {\em Remapping Period (RP)} is the total time taken to remap the v-group. With RR=1\% and two million rows in memory, a v-group will have a remap-period of about 200 million activations to the v-group.  We can reduce the remapping-period  by dividing the v-group, such that every Nth row in the v-group form a {\em v-segment} and each v-segment has an independent set of keys and Ptr.  With N=32, the remapping-period of the v-segement would be 6.25 million activations, however, this  requires 16 KB SRAM overhead for metadata.

\begin{figure*}
   \vspace{-2mm}
    \centering
    \includegraphics[width=2.05\columnwidth]{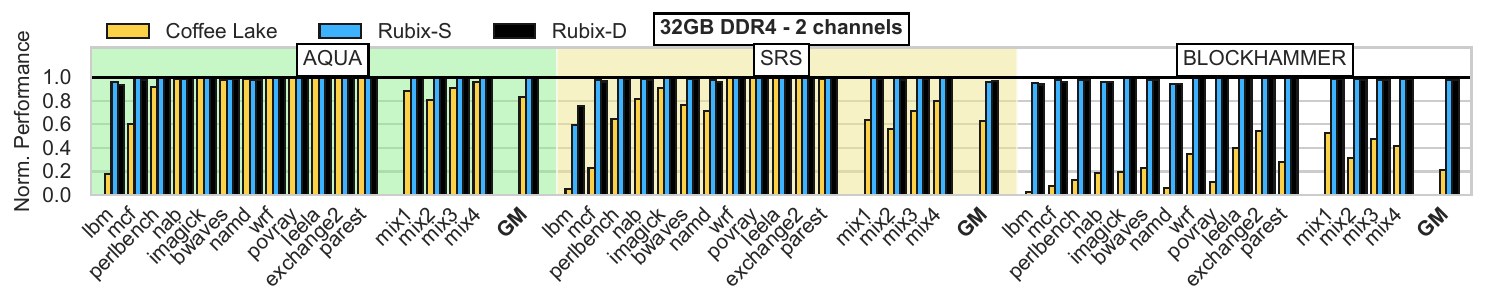}
   \vspace{-3mm}
     \includegraphics[width=2.05\columnwidth]{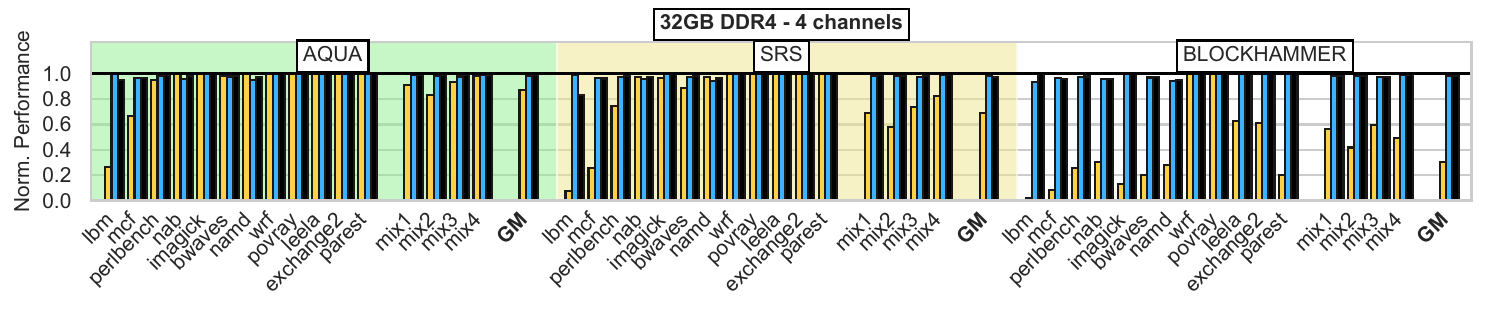}     
         \label{fig:start_t16}
    \caption{\revhl{Performance of secure mitigations with Intel and Rubix mappings, normalized to an unprotected baseline, for an 8-core multi-channel system. 
    While Intel mappings incur impractical average overheads of 15\%-380\% (AQUA-BlockHammer), Rubix reduces it to 1\%-4\%.
    }}
    \label{fig:multi_channel}
\end{figure*}

\subsection{Results: Impact on Mitigations}

\begin{figure}[!htb]
    \centering
    \includegraphics[width=1\columnwidth]{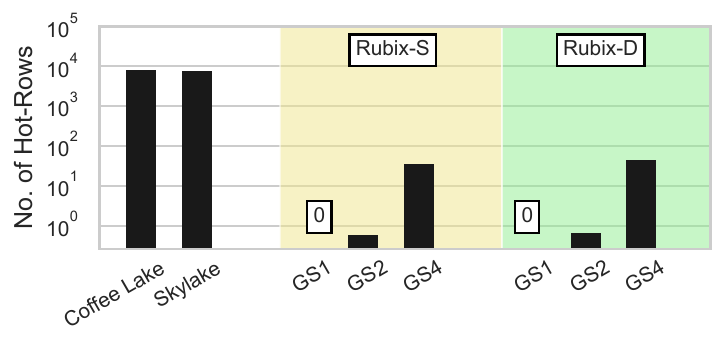}
   \vspace{-0.2in}
    \caption{Hot-rows in baseline and Rubix (100x-10000x fewer).}
    \vspace{-0.1in}
    \label{fig:rubix_mitigations}
\end{figure}

Rubix-D reduces the number of hot-rows within 64ms as shown in \cref{fig:rubix_mitigations}, which plots hot-rows for conventional policies, Rubix-S, and Rubix-D (as GS is varied). The baseline policies each have more than 7K hot-rows. Rubix with GS1 eliminates hot-rows which GS2 incurs a negligible number of hot-rows, which increase to few tens with GS1. The reduction in hot-rows makes secure mitigations viable at  $T_{RH}$ of 128. 

\subsection{Results: Impact on Performance}

We evaluate Rubix-D with Remapping-Rate of 1\% without any segments as they do not impact performance (they affect the Remapping-Period and storage overheads). \cref{fig:rubix_d_perf} shows the performance of Rubix-D compared to Intel mappings, normalized to an unprotected Coffee Lake baseline. Rubix-D incurs low overhead of just 1-3\% on average at $T_{RH}$ of 128. 
AQUA, SRS, and BlockHammer perform best at different gang-sizes. 
AQUA launches almost no mitigations and benefits from row buffer locality at GS4.
SRS operates at a lower threshold of $T_{RH}/3$ and launches more mitigations, performing best at GS2 with negligible hot rows. 
BlockHammer has high mitigation overhead and works best minimal hot-rows at GS1. 
Rubix-D incurs worst-case slowdown of just 10\%, compared to more than 100X in baseline (for BlockHammer). 
The remapping of Rubix-D also avoids getting stuck with an accidentally bad mapping,  as the mapping gets changed over program execution. 

\subsection{Results: Storage and Power Overheads}

Rubix-D needs 8-byte metadata (currKey, nextKey, Ptr) for each v-group, so 512 bytes for gang-size of 4 lines. For segmented Rubix-D, the storage overhead is proportional to the number of segments (e.g. 16KB  SRAM for 32 segments). DRAM power, computed using Micron's power calculator~\cite{micron:calc}, increases by 130mW at GS4 (4.2\% more than baseline), 180mW at GS2 (5.8\% increase), and 320mW at GS1 (10.9\% increase). 
We note that with the baseline  mapping, secure mitigation schemes not only incur significant slowdowns but also energy overheads.

\subsection{\revhl{Results: Scaled-up Multi-Channel Systems}}

We evaluate Intel Coffee Lake and Rubix mappings on a subset of workloads with 8-core simulations with 2 and 4 channels (32GB DDR4 memory and 16 MB LLC, other configuration same as in \cref{table:system_config}). As \cref{fig:multi_channel} shows, Intel's mapping incurs impractical overheads of 15\%, 45\%, and 380\% for AQUA, SRS, and BlockHammer (bottom graph), even though it stripes gangs of 4 lines across 4 channels, because contiguous lines end up in the same row in a strided pattern. Rubix breaks the spatial correlation of line-to-row, resulting in low overheads of just 1-3\% (4\% for 2-ch SRS with Rubix-S).

\subsection{Security Analysis of Rubix-D}

\revhl{Even though Rubix-D remaps dynamically, it is not a standalone mitigation for Rowhammer, as an adversary can use Flush+Reload~\cite{yarom2014flush+} to cause bit-flips. Thus, Rubix-D must always be used with a  Rowhammer mitigation scheme.} 
Rubix-D's security stems from the underlying mitigation (AQUA/SRS/Blockhammer). As the security of these schemes is not dependent on line-to-row mapping, Rubix-D retains their security (please see Section~\ref{sec:rubixs-security}).  Thus, per Lemma-1 and the fact that Rubix-D is simply a memory mapping, the overall design (with AQUA,SRS, Blockhammer) of Rubix-D is secure against all access patterns, including Half-Double.

\subsection{Impact of Rubix-D on Future Attacks}

Complex attacks, such as Half-Double and BLASTER~\cite{lang2023blaster}, work by using multiple rows to cause a bitflip.  A key ingredient of such attacks is to identify the neighbors of a given row~\cite{cojocar2020we}.  With Rubix-D, not only do we get security for known attacks, it would make orchestrating future complex pattern attacks much harder. For example, we estimate that it would take in the order of days to encounter a pair of neighboring rows for a given target row.  With Rubix-S this information remains valid till the next reboot of the system, whereas with Rubix-D this neighbor information gets changed in less than 1 second due to remapping. Thus, Rubix-D will make it much harder to orchestrate future complex pattern attacks that rely on spreading activations on multiple neighboring rows.  

Table~\ref{table:rubix-d} shows the average time to encounter a pair of neighboring rows for a given target row for Rubix-S and Rubix-D (32 segments and a rate of remapping of 1\%) and the time during which this information remains valid. Rubix-S uses static mapping, so even though the mapping is randomized, this mapping remains constant until the system reboot.  With Rubix-D, the time to encounter neighbor rows increases from 27.5 hours to about 38 years, and this mapping gets changed within the next 1 second. For this simplified analysis, we assume the attacker launches an activation every nanosecond to the bank.

\begin{table}[!htb]
  \centering
 \vspace{-0.15 in}
  \begin{small}
  \caption{Difficulty in conducting targetted attacks with Rubix-D}
  \label{table:rubix-d}
  \setlength{\tabcolsep}{3pt}
  \renewcommand{\arraystretch}{1.15}
  \begin{tabular}{|c||c|c|}
    \hline
    \textbf{Metric} & \textbf{Rubix-S} & \textbf{Rubix-D}\\ \hline \hline

    Time to encounter neighbor rows & 27.5 hours &  38.5 years  \\ \hline
    Time mapping remains valid & Until reboot & $<1$ second \\ \hline
  \end{tabular}
  \end{small}
  \vspace{-0.15 in}
\end{table}

%% file: sections/6.related.tex
\section{Related Works}
\label{sec:related}

\subsection{Mapping of Memory Systems}

\begin{figure}[!htb]
    \centering
    \includegraphics[width=1\columnwidth]{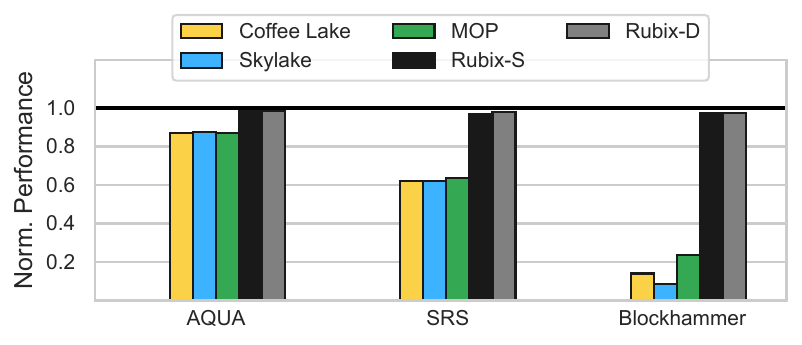}
   \vspace{-0.3 in}
    \caption{Performance of AQUA, SRS, and Blockhammer on MOP and Rubix. MOP suffers large slowdowns.}
    \label{fig:mop}
   \vspace{-0.05 in}
\end{figure}

{\em Minimalist Open-Page (MOP)}~\cite{mop} balances both performance and fairness by placing only four lines of a 4KB page in the same row. Unfortunately, as MOP round-robins across all banks, spatially proximate lines from consecutive pages are co-resident in the same row.
This maintains spatial correlation and we find hot-rows with MOP is similar to our baseline mapping. Figure~\ref{fig:mop} shows the normalized performance of AQUA, SRS, and Blockhammer for MOP, Rubix-S and Rubix-D. We observe that MOP still suffers significant slowdowns, whereas Rubix virtually eliminates the hot-rows and the associated slowdown. Rather than hand-crafting a mapping, our work uses encryption for breaking the spatial correlation of lines. 

Coffee Lake and Skylake~\cite{wang2020dramdig} contain xor functions for selecting the bank. 
However, such bank-selection functions do not change lines that co-reside in a given row.
As such, these policies do not reduce the number of hot-rows.

\subsection{Randomization in Memory Systems}

Randomization is a popular technique to improve the reliability and security of memory systems.  For example, Start-Gap~\cite{start-gap} and Security-Refresh~\cite{securityrefresh} randomize mapping in non-volatile memories for wear-leveing.
Cache randomization~\cite{ceaser,ceaser-s,mirage,scattercache,RPCache,NewCache} techniques randomize the line-to-set mapping to mitigate conflict-based cache attacks. 
 
\subsection{\revhl{In-DRAM Rowhammer Mitigations}}

DRAM modules contain {\em Target Row Refresh (TRR)}, which tracks aggressors and refreshes victims. Recent attacks~\cite{frigo2020trrespass, jattke2021blacksmith}, break TRR by exploiting its insufficient tracking capability.
Samsung's DSAC~\cite{samsung_dsac} and SK Hynix's PAT~\cite{isscc23} improve TRR for DDR5, but due to severe area limitation in DRAM, still allow aggressors to escape detection. 
DSAC has an escape probability of 13.9\% between two mitigations and PAT fails 6.9\% of the time (compared to DDR4-TRR). 
Two recent whitepapers from JEDEC\cite{JEDEC-RH1, JEDEC-RH2} mention that ``in-DRAM mitigations cannot eliminate all forms of Rowhammer attacks".  

Even if all aggressors are tracked accurately, victim-refresh is not secure as it preserves spatial proximity between aggressor and victims, enabling attacks such as Half-Double. 
Instead, our solution Rubix makes secure Rowhammer mitigations resilient to complex attacks practical at ultra-low thresholds, as shown in ~\cref{table:mitigation_compare}. As Rubix is a memory mapping, it is compatible with any tracking and mitigation mechanism.
Note that Rubix will also greatly reduce the overheads of RFM-friendly mitigations (~\cite{shadow, marazzi2023rega}) by eliminating root cause of overheads – hot-rows, thereby requiring much less RFM commands. 

\begin{table}[!htb]
  \centering
 \vspace{-0.15 in}
  \begin{small}
  \caption{\revhl{Comparison of Rowhammer Mitigations}}
  \label{table:mitigation_compare}
  \setlength{\tabcolsep}{3pt}
  \renewcommand{\arraystretch}{1.15}
  \begin{tabular}{|c||c|c|}
    \hline
    \textbf{Mitigation} & \textbf{Security} & \textbf{Slowdown}\\ \hline \hline
    in-DRAM TRR  & {\bf Not Secure} & $<2\%$ \\ \hline
    AQUA & {\bf Secure} -- Isolation & 15\% \\ \hline
    SRS & {\bf Secure} -- Randomization & 60\% \\ \hline
    BlockHammer & {\bf Secure} -- Rate Control & 600\% \\ \hline
    Rubix with AQUA/ & {\bf Secure} -- & 1\% to 3\% \\
    SRS/ BlockHammer &  underlying mitigation &  \\ \hline
  \end{tabular}
  \end{small}
  \vspace{-0.05 in}
\end{table}

\subsection{Randomization to Mitigate Rowhammer}

Recent row migration proposals, such as RRS~\cite{rrs}, SRS ~\cite{srs}, AQUA~\cite{aqua}, and Shadow~\cite{shadow}, mitigate Rowhammer by moving an aggressor row to another row in memory.  However, such row-to-row randomization does not change the set of lines that co-reside in the row
. Therefore, these schemes do not reduce the hot-rows in memory. Unlike these solutions, our work focuses on randomizing the line-to-row mapping.  

%% file: sections/7.conclusion.tex
\section{Conclusion}

Rowhammer gets worse as thresholds drop and attacks develop complex patterns like Half Double that defeat the commonly used victim-refresh.
Mitigations resilient to complex attacks, like AQUA, SRS, and Blockhmmer, suffer from drastic slowdown at low thresholds due to many \textit{hot-rows}.
We identify the line-to-row mapping as the root cause of hot-rows, as it places spatially correlated lines in same row.
Our proposal, Rubix, breaks this spatial correlation by randomizing the line-to-row mapping, reducing the number of hot rows by more than 100x. Rubix reduces overheads of the prior schemes by 10-100x, making them viable for practical adoption.